\documentclass[aps,nofootinbib,showpacs,preprintnumbers,amsmath,amssymb]{revtex4}
\usepackage{epsfig}
\usepackage{epsf}
\usepackage{amssymb}
\usepackage{amsmath}

\begin{document}
\preprint{USM-TH-219}

\title{Exponentially Modified QCD Coupling}

\author{Gorazd Cveti\v{c}$^{a,b}$}
  \email{gorazd.cvetic@usm.cl}
\author{Cristi\'an Valenzuela$^a$}
  \email{cristian.valenzuela@usm.cl}
\affiliation{$^a$\,Dept.~of Physics, Universidad T\'ecnica
Federico Santa Mar\'{\i}a, Valpara\'{\i}so, Chile\\
$^b$\,Center of Subatomic Studies, UTFSM,
Valpara\'{\i}so, Chile}

\begin{abstract}
We present a specific class of models for an infrared-finite analytic QCD coupling,
such that at large space-like energy scales the coupling differs from the perturbative one by less than any inverse power of the energy scale.
This condition is motivated by the ITEP Operator Product Expansion philosophy.
Allowed by the ambiguity in the analytization of the perturbative coupling,
the proposed class of couplings has three parameters.
In the intermediate energy region,
the proposed coupling has low loop-level and renormalization scheme dependence.
The present modification of perturbative QCD must be considered as a phenomenological attempt,
with the aim of enlarging the applicability range of the theory
of the strong interactions at low energies.

\end{abstract}

\maketitle

\section{Introduction}

The construction of an analytic coupling  $\mathcal{A}_1(Q^2)$,
a coupling analytic in the $Q^2$-plane excluding the negative (time-like) semiaxis,
is a manner to provide the evaluated observables with the analytic properties required by causality.
Perturbative QCD (pQCD), based on a
truncated power series of the $\beta$-function,
yields a coupling $a_{\rm pt}(Q^2) \equiv \alpha_s(Q^2)/\pi$
which contains (Landau) singularities in the
low energy space-like regime ($0 < Q^2 \leq \Lambda^2$).
These singularities are problematic as they result in
nonphysical singularities of the space-like
QCD observables ${\cal D}(Q^2)$ when the latter are
evaluated as truncated perturbation series (TPS)
in the pQCD coupling $a_{\rm pt}(\mu^2)$
(with renormalization scale $\mu^2 = \kappa Q^2$, $\kappa \sim 1$).
The most direct construction of an analytic coupling
was performed by Shirkov and Solovtsov \cite{ShS}
(minimal analytic coupling -- MA; for applications, see
\cite{Milton:1997mi,Sh,Shirkov:2006gv}),
who kept the discontinuities of $a_{\rm pt}(Q^2)$
unchanged on the time-like axis, but removed them
from the space-like axis.\footnote{
Recent contributions within the approach of Milton and Solovtsov are
to express higher order perturbative couplings as a series expansion in powers of the exact two-loop coupling,
and to study properties of the corresponding analytic couplings \cite{Kurashev:2003pt}; and
to analytize noninteger power of the running coupling
and explore its applications \cite{Bakulev:2006ex,Broadhurst:frac}.}
Other models for an analytic coupling
have been proposed afterwards
\cite{Nesterenko,Sanda:1979xp,Cvetic:2006ri,Cvetic:2006mk,Cvetic:2006gc},
which change in general the low
energy behavior with respect to the MA coupling.
For a review of various models, see \cite{Prosperi:2006hx}. \\

At high energies $Q^2 > \Lambda^2$,
all these couplings differ from $a_{\rm pt}(Q^2)$
by $\sim (\Lambda^2/Q^2)$. In such models the
power suppressed terms in space-like QCD observables would
come, at least partly, from the ultraviolet (UV) regime.
This would contradict
the philosophy of the ITEP Operator Product Expansion
(ITEP-OPE) approach \cite{Shifman:1978bx}.
If we want the ITEP-OPE approach to survive
in analytic QCD models, then
$\delta \mathcal{A}_1(Q^2) \equiv \mathcal{A}_1(Q^2) - a_{\rm pt}(Q^2)$
at large $Q^2$ ($\gg \Lambda^2$) must fall faster than any inverse power of $Q^2$.
We will show that,
within the context of one-chain resummations,
even in the case when $\delta \mathcal{A}_1(Q^2)$ at $Q^2 \gg \Lambda^2$ is a very suppressed {\em power} correction
$\delta \mathcal{A}_1(Q^2)\sim (\Lambda^2/Q^2)^{k_{\rm max}}$
($k_{\rm max}\gg 1$),
the UV regime still contributes to the space-like observable $\mathcal{D}(Q^2)$ an appreciable power
$\sim (\Lambda^2/Q^2)^n$.
Here $z=n$ is the location of the leading IR renormalon of $\mathcal{D}$ in the Borel plane.
On the other hand, such term has the same power-behavior as the leading OPE term of the observable, which is of IR nature.
This case would thus contradict the ITEP-OPE philosophy.
In the present paper we shall study this case in detail
and we shall see that the definition of UV and IR contributions in the renormalon-like resummation
is a crucial point.  \\

In Ref.~\cite{Alekseev}, a coupling
with $|\delta \mathcal{A}_1(Q^2)| \leq {\cal O}( (\Lambda^2/Q^2)^3)$
is constructed, by adding two specific power-suppressed
terms to the MA-coupling.
In Ref.~\cite{Raczka},
a class of analytic couplings is obtained from modified
$\beta$-functions.
The obtained
$\delta \mathcal{A}_1(Q^2)$ is at large $Q^2$ smaller than any inverse
power of $Q^2$.
Both types of couplings \cite{Alekseev,Raczka} are divergent at $Q^2 \to 0$. \\

In this work we construct a class of infrared-finite analytic couplings $\mathcal{A}_1(Q^2)$.
The proposed couplings deviate from the perturbative one by terms exponentially suppressed at high $Q^2$,
thus they differ from $a_{\rm pt}(Q^2)$ by less than any inverse power of $Q^2$.
While in the UV region $\mathcal{A}_1(Q^2)$ mimics the perturbative coupling,
the IR behavior of  $\mathcal{A}_1(Q^2)$ is modeled by means of a set of parameters (three).
In the intermediate energy region,
the proposed coupling has low loop-level and RS dependence.
The reason why the coupling is modeled lies, on one hand,
in the fact that there is not a unique way to analytize the perturbative coupling and,
on the other hand,
in the possibility of enlarging the applicability range of pQCD
by maximizing the description power of the theory at low energies.
This must be considered a phenomenological attempt where the relevant
energy region is $Q\approx 1-2$ $GeV$.
 \\

In Section 2, we present the long-distance and the
short-distance sources of the power corrections
$(\Lambda^2/Q^2)^n$ at the level of the
one-chain resummations (the leading skeleton, or
equivalently, the large-$\beta_0$ approximation).
In Section 3, we present the construction of a
class of such infrared-finite space-like analytic couplings
$\mathcal{A}_1(Q^2)$ which differ from $a_{\rm pt}(Q^2)$
by less than any power $(\Lambda^2/Q^2)^n$ for large $Q^2$.
We also present the associated Minkowskian coupling.
In Section 4,
we discuss the loop-level dependence and renormalization scheme (RS) fixing
of low-energy observables.
We also review the method of evaluation of observables
developed in our previous works \cite{Cvetic:2006mk,Cvetic:2006gc}
for general analytic QCD models,
and see how it can be applied in the
case of the here constructed analytic coupling.
Section 5 contains the conclusions of this work.

\section{Power Corrections from One-Chain Resummations}
\label{Section:PowCorr}

In this Section we consider different contributions to the one-chain resummation term $\mathcal{D}_\text{res}(Q^2)$,
paying special attention to their IR or UV character.
We are mainly interested in power-behaved contributions.\\

The one-chain resummation term is given by

\begin{equation}
\mathcal{D}_\text{res}(Q^2)=
\int_0^\infty \frac{dt}{t}  F_{\mathcal{D}}^{\cal{E}}(t) \,
\mathcal{A}_1(t\, e^C Q^2),
\label{res}
\end{equation}
where $F_{\mathcal{D}}^{\cal{E}}(t)$ is the observable-dependent Euclidean characteristic function and $\mathcal{A}_1(Q^2)$ the observable-independent running coupling,
normalized as $\alpha(Q^2)/\pi$.
(The constant $C$ is the one-loop vacuum polarization renormalization constant. It has to be included in order to make $\mathcal{D}_\text{res}(Q^2)$ $C$-independent. In the following we use the V-scheme, i.e. $C=0$, unless otherwise noted.)
We consider a running coupling, $\mathcal{A}_1(Q^2)$,
analytic in the whole $Q^2$ complex plane excluding the Minkowskian semiaxis,
i.e. a so called analytic coupling.
The analytic coupling must merge with the perturbative one in the UV limit.
In the UV expansion of  $\mathcal{A}_1(Q^2)$ one finds, in general,
in addition to the usual logarithmically varying terms of the perturbative treatment,
also power-behaved terms of the form $(1/Q^2)^n$, with $n=1,2,\ldots$.
We call them {\it running coupling power terms}.
The $t$-integration in Eq.\ (\ref{res}) is an Euclidean momentum integration,
where the running coupling is evaluated at the flowing momentum scale, $t^{1/2}Q$.
This momentum is identified with the momentum passing through a photon/gluon propagator.
In QED there is a correspondence between vacuum polarization and charge renormalization which allows the identification.
In QCD there is no such a correspondence and the identification is naive.\\

We aim to identify IR and UV contributions to $\mathcal{D}_\text{res}(Q^2)$
by comparing the momentum flowing through the boson propagator, $t^{1/2}Q$,
with the scale where QCD becomes non-perturbative.
For simplicity, we make a sharp division between IR and UV regions.
The contributions coming from the integration region where $t\, Q^2<\kappa\, \Lambda^2$ are considered to be IR,
while those coming from $t\, Q^2>\kappa\, \Lambda^2$ are considered to be UV.
The constant $\kappa$ is chosen to be bigger than 1,
and of the order or a couple of orders of magnitude bigger than 1.\footnote{
For the treatment below,
the specific value of $\kappa$ is irrelevant,
it can be chosen to be 2, 10 or 100;
of importance is its $Q^2$-independence.
The use of a sharp division between IR and UV regions is also not decisive.
The point is that, using this criterium, the minimal value of $t$ in the one-chain resummation integral Eq. (\ref{res}) from which
the contribution is considered to be UV is $Q^2$-dependent.}
Note that these IR and UV regime identifications
are different from those where the IR and UV
regimes are identified as $t<1$ and $t>1$, respectively.
If  $Q^2\sim \Lambda^2$, the IR and UV regions are similar in both IR/UV identification criteria,
but differ for $Q^2 >> \Lambda^2$. \\

In order to be concrete, we take as an example the simplest analytic coupling:
the one-loop minimal analytic coupling \cite{ShS}:

\begin{equation}
\mathcal{A}_1^{(\text{MA})}(Q^2)=
\frac{1}{\beta_0} \Big(
\frac{1}{\log Q^2/\Lambda^2}
- \frac{\Lambda^2}{Q^2-\Lambda^2} \Big),
\label{1lma}
\end{equation}
where the first term is the perturbative one-loop running coupling,
$a_\text{pt}(Q^2)\equiv\alpha_\text{pt}(Q^2)/\pi=(\beta_0\,\log Q^2/\Lambda^2)^{-1}$,
and the second term is introduced in order to make $\mathcal{A}_1^{(\text{MA})}(Q^2)$ analytic in the mentioned region without changing the discontinuity along the Minkowskian semiaxis.
We inspect now the contributions to $\mathcal{D}_\text{res}(Q^2)$ from both terms separately.

\subsection{Perturbative Coupling Contributions}

The perturbative coupling contribution to $\mathcal{D}_\text{res}(Q^2)$ is

\begin{equation}
\mathcal{D}_\text{res}^\text{I}(Q^2)=
\int_0^\infty \frac{dt}{t}  F_{\mathcal{D}}^{\cal{E}}(t) \,
\frac{1}{\beta_0\,\log t\, Q^2/\Lambda^2}.
\label{resI}
\end{equation}
We shall consider different aspects of the latter expression.
First we obtain its contributions to the usual perturbation series,
expanding the $\log^{-1}$ inside the integral.
Then, we give the usual definition of the integral ambiguity,
a reflection of the fact that the integral in Eq.\ (\ref{resI}) is not a well defined quantity.
Finally, we consider the contribution from the IR integration region. \\

In order to obtain the part of the perturbation series which is contained in Eq.\ (\ref{resI}),
we expand the running coupling
$a_\text{pt}(t\, Q^2)$ around
a chosen renormalization scale $t_*\,  Q^2$:

\begin{equation}
a_\text{pt}(t\, Q^2)=
\sum_{n=0}^\infty (-\beta_0 \log t/t_*)^n\,
a_\text{pt}^{n+1}(t_* Q^2).
\end{equation}
Replacing in Eq.\ (\ref{resI}) and exchanging the order of integration and summation, we obtain

\begin{equation}
\mathcal{D}_\text{res}^\text{I}(Q^2)\longrightarrow
\sum_{n=0}^\infty \beta_0^n f_n(t_*)
a_\text{pt}^{n+1}(t_* Q^2),
\label{leadb0}
\end{equation}
where

\begin{equation}
f_n(t_*)=
\int_0^\infty \frac{dt}{t}  F_{\mathcal{D}}^{\cal{E}}(t)
(-\log t/t_*)^n.
\end{equation}
The series obtained in Eq.\ (\ref{leadb0}) is the large-$\beta_0$ expansion of $\mathcal{D}(Q^2)$,
which consists of all the terms of the perturbation series of the form $\beta_0^n a_\text{pt}^{n+1}$.
If we had considered in Eq.\ (\ref{resI}) another perturbative running coupling, solution of a higher order renormalization group equation,
we would have obtained some further terms (beyond the large-$\beta_0$) of the perturbation series of the observable $\mathcal{D}(Q^2)$.\\

The integral in  Eq.\ (\ref{resI}) is not well defined due to the simple pole at $t=\Lambda^2/Q^2$.
Clearly, replacing $a_\text{pt}(Q^2)$ by an analytic coupling $\mathcal{A}_1(Q^2)$ one gets a  well defined quantity as our starting point, Eq.\ (\ref{res}), is.
In order to give a meaning to $\mathcal{D}_\text{res}^\text{I}(Q^2)$ one chooses a modification of the integration contour,
relevant only at the pole.
The most used prescription is to take the principal value of Eq.\ (\ref{leadb0});
the procedure, however, yields a non-analytic but piecewise analytic contribution,
where the non-analytic term is purely imaginary \cite{Caprini:2007vn}.
Related to the prescription choice there is an ambiguity,
the standard definition of the ambiguity is:
the difference between
Eq.\ (\ref{resI}) with the integration contour slightly above the real axis and
Eq.\ (\ref{resI}) with the integration contour slightly below the real axis.
As a result we get:

\begin{equation}
Amb^\text{std}[\mathcal{D}_\text{res}^\text{I}(Q^2)]=
-\frac{2\pi i}{\beta_0}\, F_{\mathcal{D}}^{\cal{E}}(\Lambda^2/Q^2).
\label{AmbStd1}
\end{equation}
In the following we consider the small $t$ behavior of $F_{\mathcal{D}}^{\cal{E}}(t)$ to be given by

\begin{equation}
F_{\mathcal{D}}^{\cal{E}}(t)\approx k\, t^n, \qquad \text{for }t\ll 1.
\label{appF}
\end{equation}
Thus, for $Q^2\gg \Lambda^2$ one gets

\begin{equation}
Amb^\text{std}[\mathcal{D}_\text{res}^\text{I}(Q^2)]=
-\frac{2\pi i}{\beta_0}\, k\,
\Big(\frac{\Lambda^2}{Q^2}\Big)^n +\ldots.
\label{AmbStd2}
\end{equation}
The ambiguity Eq.\ (\ref{AmbStd1}) contains the contributions of all large-$\beta_0$ IR renormalons.
These poles lie on the Borel plane at $u=n$ where $n$ is a positive integer \cite{Beneke:1998ui}.
For the relation between the representations of the  large-$\beta_0$ contribution, using a one-chain resummation or using a Borel integral,
see \cite{Brooks:2006it}.
In the present formulation,
the ambiguity (\ref{AmbStd1}) depends on the characteristic function and on the (non)analyticity properties of the perturbative running coupling
-- Landau singularity (pole at $t Q^2 = \Lambda$) \cite{Neubert:1994vb}.
For inclusive observables, such as sum rules, the ambiguity of the type
(\ref{AmbStd1}) must get canceled by the ambiguity of the corresponding power
term in the Operator Product Expansion  (OPE)  \cite{David:1983gz,David:1985xj}. \\

The contribution from the IR integration region,
$t< \kappa\, \Lambda^2/Q^2$, is given by

\begin{equation}
\int_0^{\kappa\, \Lambda^2/Q^2} \frac{dt}{t}  F_{\mathcal{D}}^{\cal{E}}(t) \,
\frac{1}{\beta_0\,\log t\, Q^2/\Lambda^2}.
\label{PtLSIR}
\end{equation}
For $Q^2\gg\Lambda^2$ its leading term is $\sim (\Lambda^2/Q^2)^n$.
The latter integration region includes the position of the running coupling pole and
Eq.\ (\ref{PtLSIR}) has the same leading $Q^2$-dependence as the ambiguity.

\subsection{Contributions from Power Terms}

The difference between the $\mathcal{D}_\text{res}(Q^2)$ and the perturbative coupling contribution is given,
in the case of the one-loop minimal analytical coupling, by

\begin{equation}
\mathcal{D}_\text{res}^\text{II}(Q^2)=
\mathcal{D}_\text{res}(Q^2)-\mathcal{D}_\text{res}^\text{I}(Q^2)=
\int_0^\infty \frac{dt}{t}  F_{\mathcal{D}}^{\cal{E}}(t) \,
\frac{1}{\beta_0}
\Big(
\frac{-\Lambda^2}{t\,Q^2 -\Lambda^2}
\Big).
\label{resII}
\end{equation}
The integrand has a simple pole at the same $t$-value as the integrand in Eq.\ (\ref{resI}).
The corresponding ambiguity,
$+(2\pi i/\beta_0) F_{\mathcal{D}}^{\cal{E}}(\Lambda^2/Q^2)$,
has equal absolute value and opposite sign as the one of $\mathcal{D}_\text{res}^\text{I}(Q^2)$,
therefore cancels in
$\mathcal{D}_\text{res}=\mathcal{D}_\text{res}^\text{I}+\mathcal{D}_\text{res}^\text{II}$.
This implies that, if OPE is applied in the analytic approach,
the OPE terms must also have their ambiguity lifted.
One can also see that the IR part of Eq.\ (\ref{resII}) has the same leading $Q^2$-dependence  $(\Lambda^2/Q^2)^n$ as the IR part of $\mathcal{D}_\text{res}^\text{I}(Q^2)$,
for $Q^2\gg \Lambda^2$.\\

Now we turn to the aspect we are mainly interested in, the UV contribution:

\begin{equation}
\int_{\kappa\, \Lambda^2/Q^2}^\infty \frac{dt}{t}  F_{\mathcal{D}}^{\cal{E}}(t) \,
\frac{1}{\beta_0}
\Big(
\frac{-\Lambda^2}{t\,Q^2 -\Lambda^2}
\Big).
\label{resIIUV}
\end{equation}
Since $\kappa>1$, inside the integral we are allowed to expand:

\begin{equation}
\frac{-\Lambda^2}{t\,Q^2 -\Lambda^2}=
-\sum_{i=1}^\infty \Big(\frac{\Lambda^2}{t\,Q^2}\Big)^i.
\end{equation}
We consider the contribution to $\mathcal{D}_\text{res}^\text{II}(Q^2)$ from each term of the latter sum individually:

\begin{equation}
-\frac{1}{\beta_0}
\int_{\kappa\, \Lambda^2/Q^2}^\infty \frac{dt}{t}  F_{\mathcal{D}}^{\cal{E}}(t) \,
\Big(\frac{1}{t}\Big)^i
\Big(\frac{\Lambda^2}{Q^2}\Big)^i.
\label{UVpow}
\end{equation}
Then, we ask ourselves what is the $Q^2$-dependence of the latter expression for $Q^2\gg \Lambda^2$.
In order to use the approximated expression for $F_{\mathcal{D}}^{\cal{E}}(t)$, for $t\ll 1$, we divide the integration interval in two parts:

\begin{equation}
\int_{\kappa\, \Lambda^2/Q^2}^\infty =
\int_{\kappa\, \Lambda^2/Q^2}^{\bar{t}} +
\int_{\bar{t}}^\infty,
\label{UVpow2}
\end{equation}
choosing (fixing) ${\bar t} <  1$ such that in the first integral the approximation for $F_{\mathcal{D}}^{\cal{E}}(t)$ given in Eq.\ (\ref{appF}) is valid.
We see that the second integration interval generates a power correction term of the form $(\Lambda^2/Q^2)^{i}$,
while the contribution from the first integration interval can be written as:

\begin{equation}\label{}
-\frac{1}{\beta_0}
\int_{\kappa\, \Lambda^2/Q^2}^{\bar{t}} \frac{dt}{t}  (k t^n) \,
\Big(\frac{1}{t}\Big)^i
\Big(\frac{\Lambda^2}{Q^2}\Big)^i +
\ldots
\sim
\begin{cases}
(\Lambda^2/Q^2)^i \log(Q^2/\Lambda^2) + \ldots   & \text{if   $i=n$,}   \\ \\
(\Lambda^2/Q^2)^{\min(i,n)}  + \ldots            & \text{if $i\neq n$.}
\end{cases}
\end{equation}
Thus, the term (\ref{UVpow}) has a $Q^2$ behavior given by

\begin{equation}\label{}
-\frac{1}{\beta_0}
\int_{\kappa\, \Lambda^2/Q^2}^\infty \frac{dt}{t}  F_{\mathcal{D}}^{\cal{E}}(t) \,
\Big(\frac{1}{t}\Big)^i
\Big(\frac{\Lambda^2}{Q^2}\Big)^i +
\ldots
\sim
\begin{cases}
(\Lambda^2/Q^2)^i \log(Q^2/\Lambda^2) + \ldots   & \text{if   $i=n$,}   \\ \\
(\Lambda^2/Q^2)^{\min(i,n)}  + \ldots            & \text{if $i\neq n$.}
\end{cases}
\end{equation}

We conclude that terms of the type $(1/Q^2)^{i}$ from the high momentum expansion of an analytic coupling $\mathcal{A}_1(Q^2)$,
introduce in the UV integration region of the one-chain resummation integral, Eq.\ (\ref{res}),
observable power corrections whose leading term is proportional to $(\Lambda^2/Q^2)^i \log(Q^2/\Lambda^2)$, if $i=n$,
and to $(\Lambda^2/Q^2)^{\min(i,n)}$, if $i\neq n$, for  $Q^2\gg \Lambda^2$. \\

Returning to expression (\ref{resIIUV}),
i.e. the UV part of $\mathcal{D}_\text{res}^\text{II}$,
the leading power-behaved contribution comes from the term with $i=1$;
and is proportional to
$\Lambda^2/Q^2$ if $n \geq 2$, or to $(\Lambda^2/Q^2) \log(Q^2/\Lambda^2)$ if $n=1$.\\

Power corrections coming from UV degrees of freedom are in conflict \cite{Ball:1995ni,Dokshitzer:1995qm} with the OPE philosophy.
Therefore, an analytic coupling having a high momentum expansion with corrections to the perturbative running coupling falling faster than any negative power of $Q^2$ is wished.
We address this question in the next Section. \\

Before doing this, let us consider a different definition of the ambiguity in the one-chain resummation term $\mathcal{D}_\text{res}^\text{I}(Q^2)$, motivated by the fact that the perturbative running coupling $a_\text{pt}(Q^2)$ has no physical meaning in the IR region.
We can define the ambiguity of $\mathcal{D}_\text{res}^\text{I}(Q^2)$ as the contribution of the IR region, i.e. $t <\kappa\, \Lambda^2/Q^2$, with  $a_\text{pt}(Q^2)$ inside the integral replaced by a constant $a_0$:

\begin{equation}
Amb[\mathcal{D}_\text{res}^\text{I}(Q^2)]=
a_0 \int_0^{\kappa\, \Lambda^2/Q^2} \frac{dt}{t}  F_{\mathcal{D}}^{\cal{E}}(t).
\label{amb2}
\end{equation}
First, we note that the ambiguity is now a real quantity,
instead of a pure imaginary one as it is conventionally defined.
Furthermore, for $Q^2\gg\Lambda^2$, the usual ambiguity and Eq.\ (\ref{amb2}) have the same leading $Q^2$-dependence $\sim (\Lambda^2/Q^2)^n$.
The ambiguity defined as in Eq.\ (\ref{amb2}) is independent of the properties of $a_\text{pt}(Q^2)$, in particular,
is independent of the loop order the perturbative running coupling is solution of,
and independent of its analyticity properties.\\

In order to have an idea of the order of magnitude of the ambiguity Eq.\ (\ref{amb2}), we take as an example the Adler function,
$Q=1.8$ $GeV$, $\Lambda_{C=0}=0.9$ $GeV$, and $\kappa=1.5$.
We get $Amb[\mathcal{D}_\text{res}^\text{I}]=0.8$ $a_0$.
The ambiguity defined in this manner serves as a measure of the effect of modeling the running coupling on one-chain resummation contributions.

\section{Exponentially Modified Coupling}\label{Section:ExpCoup}

In this Section a new analytic coupling is proposed.
In the $Q^2\gg\Lambda^2$ region, the coupling converges to the perturbative running coupling with deviations from it falling faster than any power of the momentum.
The behavior of the coupling in the IR is described through few parameters.

\subsection{Logarithm Replacement}

We start considering the one loop case.
In Section \ref{Section:PowCorr} we saw a possible replacement of the logarithm which yields an analytic coupling:

\begin{equation}
\frac{1}{\log x} \longrightarrow
\frac{1}{\log x} + \frac{1}{1-x},
\label{1lmaReplacement}
\end{equation}
where $x=Q^2/\Lambda^2$.
The latter analytization of the $1/(\log x)$ function is unique if one requires that:
(a) it keep the same discontinuity along the $x\leq 0$ semiaxis and
(b) it vanish for $|x|\rightarrow \infty$
(with $x$ complex and not lying on the negative semiaxis; and having the dispersion relation shown below).
As we discussed above, Eq.\ (\ref{1lmaReplacement}) introduces unwished power terms in the large-$x$ expansion of $\mathcal{D}_\text{res}(Q^2)$.
In order to modify the latter replacement,
we must abandon at least one of the previous conditions.
It seems unreasonable to give up condition (b).
If the coupling does not vanish for $|x|\rightarrow \infty$ in some region of the complex plane, sum rules and the expression for the Adler function in terms of the $R$ ratio would not hold. It is also against the intuition of asymptotic freedom.
On the other hand, there is no reason to maintain the same discontinuity on the time-like axis as in perturbation theory, hence we relax this condition.\\

The following replacement is an example of modification of Eq.\ (\ref{1lmaReplacement}),
that does not introduce power-behaved terms:

\begin{equation}
\frac{1}{\log x} \longrightarrow
\frac{1}{\log x} + \frac{\; e^{\nu(1-x^{a})}}{1-x},
\label{Replacement}
\end{equation}
with $\nu >0$ and $0<a\leq 1/2$.
The parameter $a$ must not exceed the value $1/2$ because otherwise condition (b) would be violated.
We define the cut of the $x^{a}$ function along the negative semiaxis as for the logarithm function.
Thus, by means of the exponential function,
we get a one-loop analytic coupling with corrections to the perturbative one falling, in the UV region, faster than any power of $x$.

\subsection{$N$-loop Coupling}\label{SubsectionNloopCoupling}

Based on the previously introduced replacement, Eq.\ (\ref{Replacement}),
we construct an analytic coupling from the $N$-loop running coupling.
We write the latter in the form:

\begin{equation}
a_\text{pt}^{(N)}=
\sum_{n=1}^N \sum_{m=0}^{n-1} k_{nm}
\frac{\log^m L}{L^n},
\label{pertcoup}
\end{equation}
where $L=\log Q^2/\Lambda^2$ and $k_{nm}$ are functions of the $\beta$-function coefficients.
The new function $\mathcal{A}_1^{(N)}(Q^2)$ should have the following properties:

\begin{enumerate}
    \item be analytic in the complex plane excluding the negative semiaxis,
    \item differ from $a_\text{pt}^{(N)}$, for $Q^2\gg\Lambda^2$, by terms falling faster than any power of $\Lambda^2/Q^2$,
    \item vanish for $|x|\rightarrow \infty$ as $a_\text{pt}^{(N)}$ (with $x\equiv Q^2/\Lambda^2$ complex and not lying on the negative semiaxis)
        and have a dispersion relation:

        \begin{equation}
            \mathcal{A}_1^{(N)}(Q^2)=
            \frac{1}{\pi} \int_0^\infty \frac{d\sigma}{\sigma + Q^2} \,\rho(\sigma),
            \label{DispRel}
        \end{equation}
            with $\rho(\sigma)= \text{Im}(\mathcal{A}_1^{(N)}(-\sigma-i \varepsilon \Lambda^2))$, and
    \item suppressed $N$-dependence (loop-level dependence) in the IR region.
\end{enumerate}
The loop-level and the renormalization scheme (choice of $\beta_2$, $\beta_3$, $\ldots$) are perturbative concepts.
Demanding the last property is necessary for a coupling largely independent of these concepts in the IR region.\\

\begin{figure}
 \centering\epsfig{file=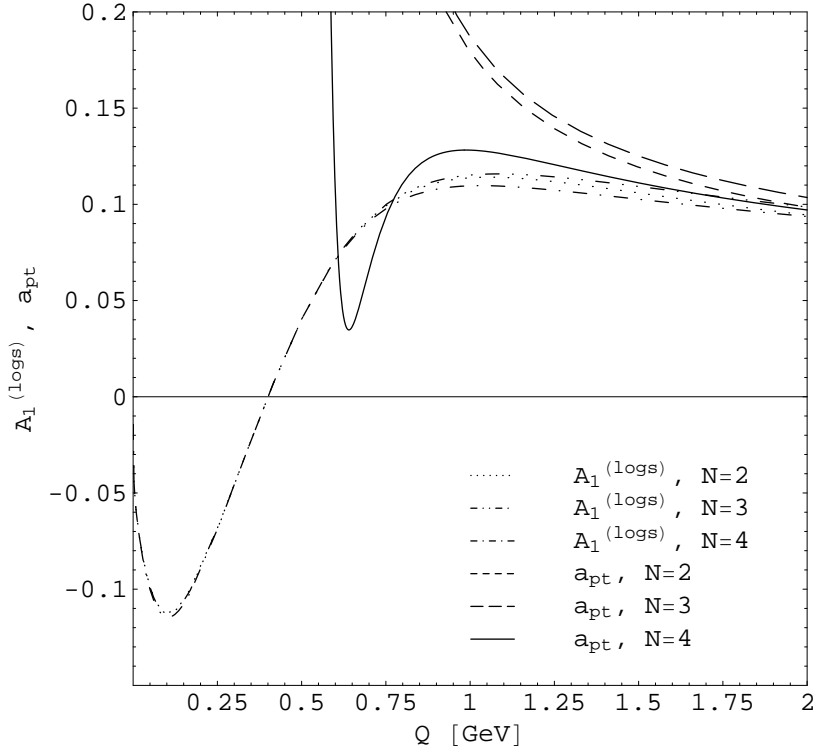}
\caption{\footnotesize The functions $\mathcal{A}_1^{(N)\text{(logs)}}$, Eq.\ (\ref{A1logs}), and $a_\text{pt}^{(N)}$, Eq.\ (\ref{pertcoup}),
are plotted as a function of $Q$, in the $\overline {\rm MS}$-scheme, with $\Lambda=0.4$  $GeV$,
for the cases $N=$ 2, 3, and 4. See comments in the text.}
\label{FigA1ll234}
\end{figure}

\begin{figure}
 \centering\epsfig{file=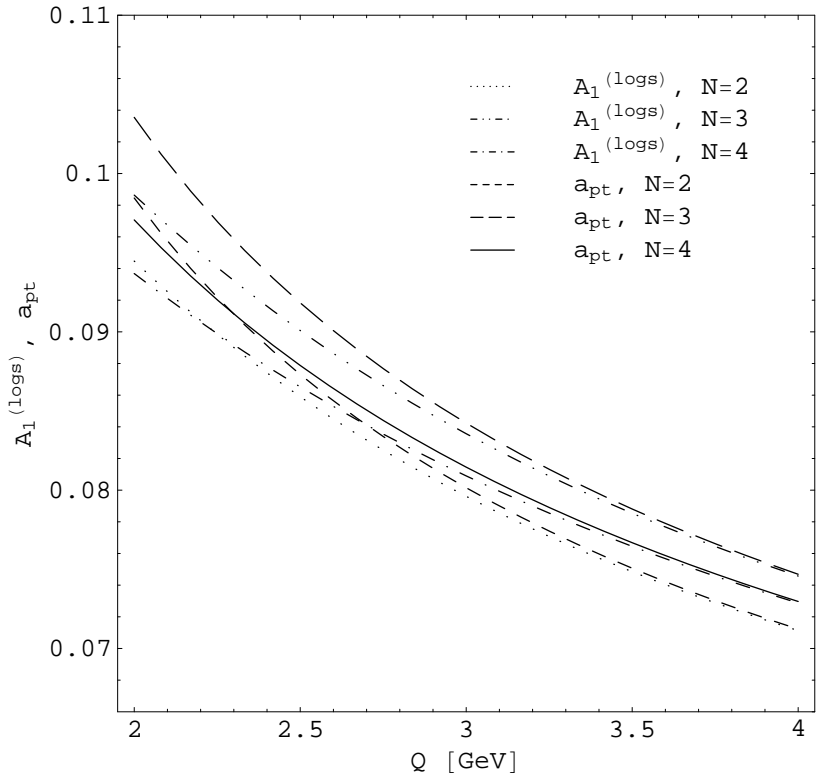}
\caption{\footnotesize Same as in Fig.\ \ref{FigA1ll234}, for a different range of energies.}
\label{FigA1ll234zoom}
\end{figure}

We propose an analytic coupling having the form

\begin{equation}
\mathcal{A}_1^{(N)}(Q^2)=
\sum_{n=1}^N \sum_{m=0}^{n-1} k_{nm}\,
\frac{\log^m L_1}{L_0^n}
+ e^{-\eta \sqrt{x}}\, f(x).
\label{coupling}
\end{equation}
The second term is only relevant in the IR region and the first term (double sum) plays, in the UV region,
the role of the perturbative coupling.
$L_0$ and $L_1$ are chosen aiming at a low $N$-dependence in the IR region.
That is achieved by suppressing $\log L_1$ and $1/L_0$ in this region.
Generalizing Eq.\ (\ref{Replacement}) we define

\begin{equation}
\frac{1}{L_i}=\frac{1}{L}+
\frac{\; e^{\nu_i(1-\sqrt{x})}}{1-x} g_i(x),  \quad \nu_i>0,  \quad i=0, 1.
\label{Ldef}
\end{equation}
The functions $g_i(x)$ and the constants $\nu_i$ are chosen as follows.
For $L_0$ we require:

\begin{enumerate}
    \item[(i)]   $1/L_0(x)$  analytic in the complex plane excluding the negative semiaxis.
                    It implies $g_0(1)=1$,
    \item[(ii)]  $1/L_0(x)$ suppressed ($\ll 1$) near $x=1$,
    \item[(iii)] $1/L_0(x)$ suppressed near $x=0$, and
    \item[(iv)]  $g_0(x)/x \leq constant$, for $|x|\rightarrow\infty$,
        in order to fulfill the previous condition 3.
\end{enumerate}
Similarly, for $L_1$ we require:

\begin{enumerate}
    \item[(i)]  $\log L_1(x)$ analytic in the complex plane excluding the negative semiaxis.
                    It implies $g_1(1)=1$,
    \item[(ii)]  $\log L_1(x)$ suppressed near $x=1$, and
    \item[(iii)] $\log L_1(x)$ suppressed near $x=0$.
\end{enumerate}
We present an analytical coupling fulfilling all previous requirements. We choose:

\begin{eqnarray}
  g_0(x) &=& \frac{2 x}{(1+\nu_0)+x(1-\nu_0)},\qquad 0<\nu_0<1; \\
  g_1(x) &=& \frac{d e^{-\nu_1}+x (d+1-d e^{-\nu_1})}{d + x},\qquad d>0.
\end{eqnarray}
We fix $\nu_0=1/2$ and $\nu_1=d=2$.
Here, $d$ is chosen such that the introduced Minkowskian pole is located in the IR region.
Thus, we have a function

\begin{equation}
\mathcal{A}_1^{(N)\text{(logs)}}(Q^2)=
\sum_{n=1}^N \sum_{m=0}^{n-1} k_{nm}\,
\frac{\log^m L_1}{L_0^n},
\label{A1logs}
\end{equation}
which depends on the $\beta$-function coefficients through $k_{nm}$ and on $N$.
In Figs.\ \ref{FigA1ll234} and \ref{FigA1ll234zoom}, $\mathcal{A}_1^{(N)\text{(logs)}}$ and the perturbative coupling $a_\text{pt}^{(N)}$ are plotted for $N=$ 2, 3, and 4, in the $\overline {\rm MS}$-scheme and for $\Lambda=0.4$  $GeV$.
In Fig.\ \ref{FigA1ll234} one can see the very low dependence of $\mathcal{A}_1^{(N)\text{(logs)}}$ on $N$,
and hence on $\beta_i$ ($i\geq 2$), in the IR region.
Fig.\ \ref{FigA1ll234zoom} shows that in the  perturbative region, for a fixed $N$, the perturbative and the analytized couplings merge.
Besides, the $N$ and $\beta_i$ dependence of  $\mathcal{A}_1^{(N)\text{(logs)}}$ is moderate in the intermediate region.

We adopt the following point of view:
the first term in Eq.\ (\ref{coupling}) gives a contribution to $\mathcal{A}_1^{(N)}(Q^2)$ in the IR region with
{\it no free parameters}
(other than $\Lambda$),
the only free parameters of the coupling being those contained in the second term of Eq.\ (\ref{coupling}).\footnote{Of course, if we had made a different choice for $L_0(x)$ and/or $L_1(x)$, the values of the free parameters in $e^{-\eta \sqrt{x}}\, f(x)$ would change.}

\begin{figure}
 \centering\epsfig{file=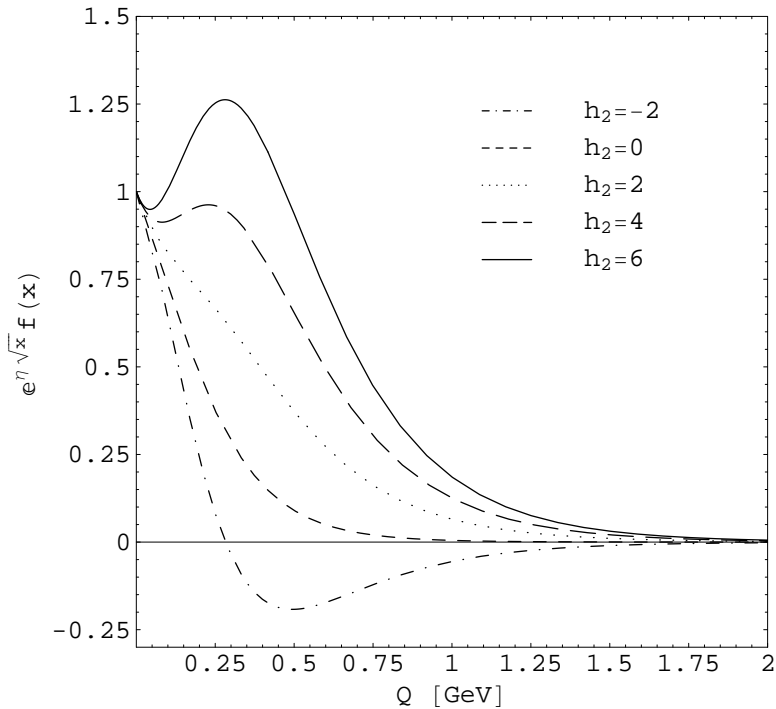}
\caption{\footnotesize The parametrized part of $\mathcal{A}_1^{(N)}$, given in Eq.\ (\ref{couplingIR}),
is plotted for $\eta=h_1=1$, $\Lambda=$ 0.4 $GeV$, and various values of $h_2$.}
\label{Figef1}
\end{figure}

\begin{figure}
 \centering\epsfig{file=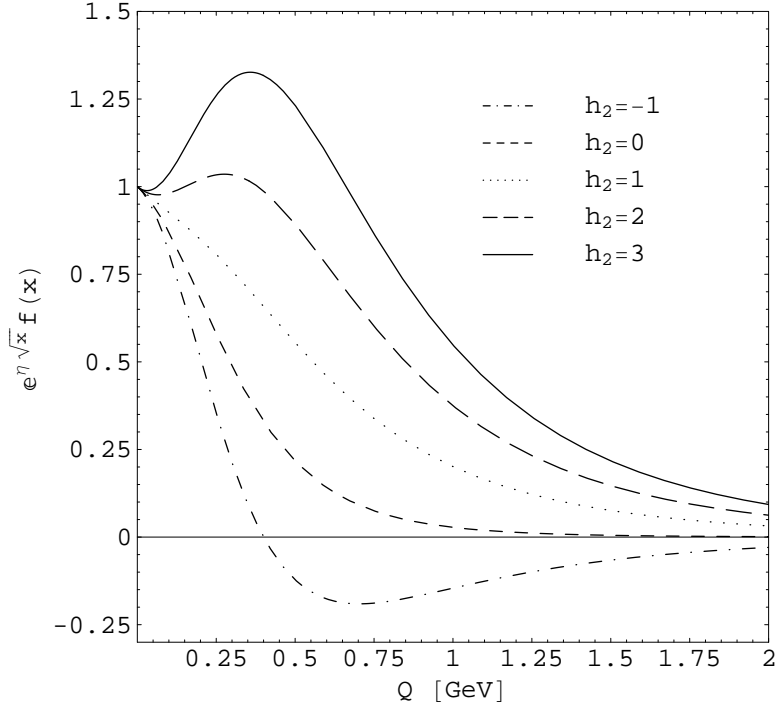}
\caption{\footnotesize Same as in Fig.\ \ref{Figef1}, with $\eta=0.3$ and $h_1=1$.}
\label{Figef2}
\end{figure}

Now we turn to the second term of Eq.\ (\ref{coupling}).
Its parametric form is chosen without a physical motivation in mind.
One just aims at a simple way of describing the coupling behavior in the IR.
A possible choice is:

\begin{equation}
e^{-\eta \sqrt{x}}\, f(x)=
h_1 \, \frac{1+h_2\, x}{(1+x/2)^2}\, e^{-\eta \sqrt{x}},
\label{couplingIR}
\end{equation}
with three parameters.
The parameter $\eta$ regulates its exponential suppression,
$h_1$ is a global factor and gives the value of this term at $x=0$, and
$h_2$ describes, to a large degree,
its behavior at $x\sim 1$.
The effect of varying $\eta$ can be observed by comparing Figures  \ref{Figef1} and \ref{Figef2}.
In each of these figures, the curves for various values of $h_2$ are plotted.

\begin{figure}
 \centering\epsfig{file=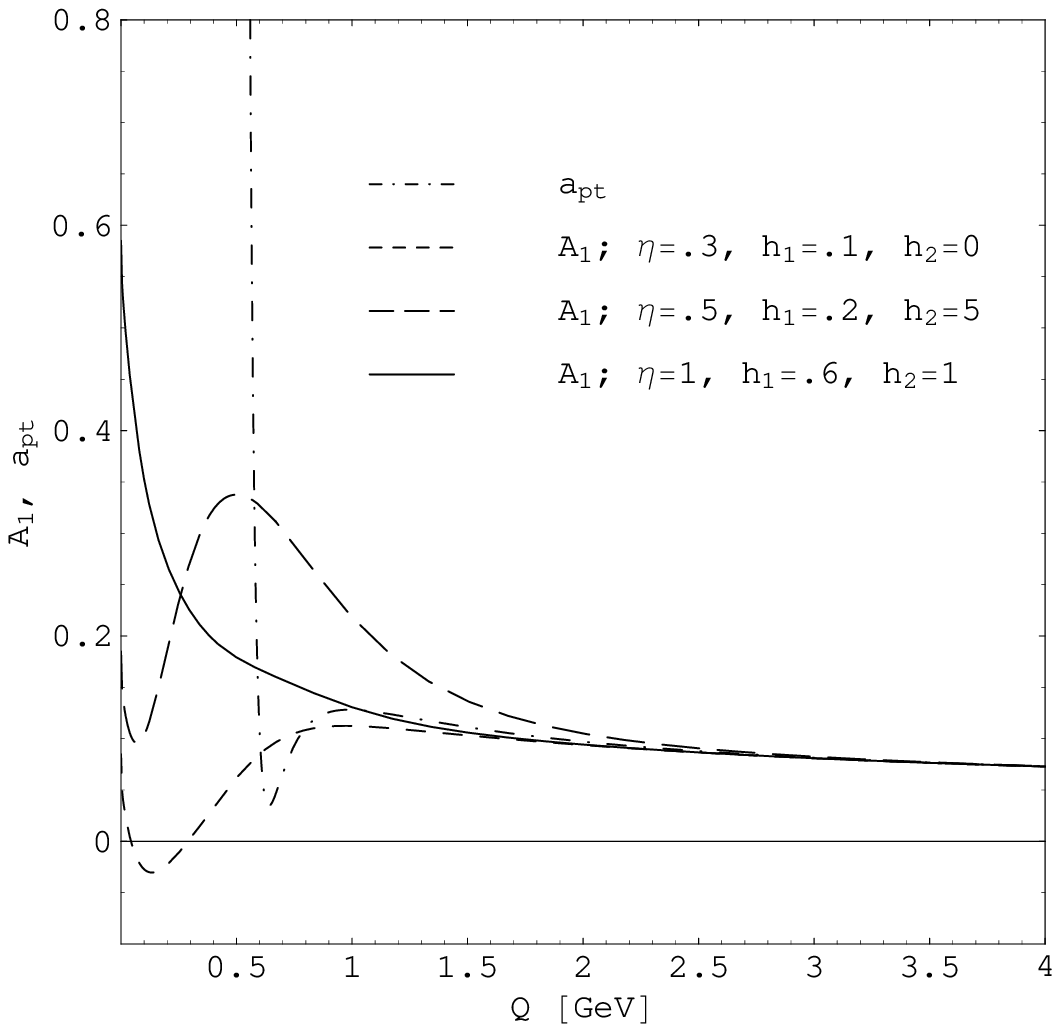}
\caption{\footnotesize Three examples of $\mathcal{A}_1^{(4)}$, Eq. (\ref{coupling}), are plotted together with the perturbative coupling $a_\text{pt}^{(4)}$.
The function $a_\text{pt}^{(4)}$ is plotted only for $Q^2> \Lambda^2$ because it has a branch cut along $Q^2\leq \Lambda^2$.
The $\overline {\rm MS}$-scheme and $\Lambda=0.4$  $GeV$ are used.}
\label{FigA1examples}
\end{figure}

In Fig.\ \ref{FigA1examples},
curves of  $\mathcal{A}_1^{(4)}$ for three different choices of parameters for are shown,
illustrating various possible IR behaviors of the analytic coupling.

The RS dependence of $\mathcal{A}_1^{(N)}$ is contained only in the coefficients $k_{nm}$,
the parametrized contribution, Eq.\ (\ref{couplingIR}), being RS independent.

\subsection{Minkowskian Coupling}

In this Subsection we present the Minkowskian coupling $\mathfrak{A}_1(s)$
corresponding to the exponentially modified one-loop coupling.
The Minkowskian (time-like) coupling  $\mathfrak{A}_1(s)$, for $s\geq 0$ is defined
from the Euclidean (space-like) coupling $\mathcal{A}_1(Q^2)$ by

\begin{equation}\label{RDrel}
\mathfrak{A}_1(s) =
\frac{-1}{2 \pi i} \int_{s-i\epsilon}^{s+i\epsilon}
\frac{dz}{z} \mathcal{A}_1(-z),
\end{equation}
where the integration path avoids the time-like semiaxis $z>0$.
The inverse relation is given by

\begin{equation}\label{DRrel}
\mathcal{A}_1(Q^2)  =  Q^2 \int_0^{\infty}
\frac{ds}{(s+Q^2)^2} \mathfrak{A}_1(s).
\end{equation}
The couplings $\mathfrak{A}_1(s)$ and $\mathcal{A}_1(Q^2)$ are related to each other
in the same manner as the $e^+ e^-$ ratio $R(s)$ and the Adler function $D(Q^2)$  are \cite{Radyushkin:1982kg,Pivovarov:1991bi}.
The coupling $\mathcal{A}_1(Q^2)$ is analytic in the complex plane excluding the Minkowskian semiaxis
and behaves appropriately at infinity
in order to have the spectral representation

\begin{equation}\label{SpectralRep}
\mathcal{A}_1(Q^2) =
\frac{1}{\pi}\int_0^\infty \frac{d\sigma}{\sigma+Q^2}\: \rho_1(\sigma),
\end{equation}
where the spectral function $\rho_1=\text{Im}[\mathcal{A}_1(-\sigma-i\varepsilon \Lambda^2)]$, with $\varepsilon\rightarrow 0$.
Replacing Eq.\ (\ref{SpectralRep}) in Eq.\ (\ref{RDrel}) the following representation for the Minkowskian coupling is obtained

\begin{equation}\label{expU1}
\mathfrak{A}_1(s) =
\frac{1}{\pi}\int_s^\infty \frac{d\sigma}{\sigma}\: \rho_1(\sigma).
\end{equation}
Returning to the evaluation of observables,
the one-chain resummation integral in Eq.\ (\ref{res}),
which involves $\mathcal{A}_1$ in the integrand,
can be written in terms of $\mathfrak{A}_1$ \cite{Dokshitzer:1995qm}

\begin{equation}
\mathcal{D}_\text{res}(Q^2)=
\int_0^\infty \frac{dt}{t}  F_{\mathcal{D}}^{\cal{M}}(t) \,
\mathfrak{A}_1(t\,Q^2),
\label{}
\end{equation}
where $F_{\mathcal{D}}^{\cal{M}}(t)$ is the observable-dependent Minkowskian characteristic function.

In this Subsection we present the spectral function and the Minkowskian coupling corresponding to the exponentially modified one-loop coupling

\begin{equation}\label{A1N1}
\mathcal{A}_1(Q^2) =
\frac{1}{\beta_0}
\Big(
\frac{1}{\log x}+
\frac{\; e^{\nu_0(1-\sqrt{x})}}{1-x}
\frac{2 x}{(1+\nu_0)+x(1-\nu_0)}
\Big)
+
h_1 \, \frac{1+h_2\, x}{(1+x/2)^2}\, e^{-\eta \sqrt{x}},
\end{equation}
with $x=Q^2/\Lambda^2$ and $\nu_0=1/2$.
The free parameters of the model are $\eta$, $h_1$, and $h_2$.
The spectral function is divided in two terms, the $N=1$ contribution and the parametrized part of the coupling (proportional to $h_1$): $\rho_1(\sigma)=\rho_1^{(N=1)}(\sigma)+\rho_1^{\text{(para)}}(\sigma)$, with

\begin{equation}\label{}
\begin{split}
\rho_1^{(N=1)}(\sigma)=
& \frac{1}{\beta_0}\frac{\pi}{(\log\sigma/\Lambda^2)^2+\pi^2} \\
& - \frac{2\sigma e^{\nu_0}}{\beta_0 (\sigma+\Lambda^2)}
\Big[
     \cos(\nu_0 \sqrt{\sigma}/\Lambda) \,\pi\, \delta((1+\nu_0)-(1-\nu_0)\sigma/\Lambda^2) \\
& \qquad \qquad
   + \sin(\nu_0 \sqrt{\sigma}/\Lambda)
        \text{Re} \Big( \frac{1}{(1+\nu_0)-(1-\nu_0)\sigma/\Lambda^2 + i \epsilon}\Big)
        \Big],
\end{split}
\end{equation}
and

\begin{equation}
\rho_1^{\text{(para)}}(\sigma)=
4 h_1 (1-h_2 \sigma/\Lambda^2)
\Big[ \pi\, \delta^{'}(\sigma/\Lambda^2 -2) \cos(\eta \sqrt{\sigma}/\Lambda)
+ \text{Re} \Big( \frac{1}{(\sigma/\Lambda^2 -2 +i \epsilon)^2}\Big) \sin(\eta \sqrt{\sigma}/\Lambda)
\Big].
\end{equation}
The Minkowskian coupling
$\mathfrak{A}_1(s)=\mathfrak{A}_1^{(N=1)}(s)+\mathfrak{A}_1^{\text{(para)}}(s)$
is obtained by replacing the spectral function in Eq.\ (\ref{expU1}).
The $N=1$ contribution is given by

\begin{equation}\label{U1N1a}
\mathfrak{A}_1^{(N=1)}(s) =
\frac{1}{\beta_0}
\Big(
-\frac{1}{2}-\frac{1}{\pi}\arctan\frac{\log(s/\Lambda^2)}{\pi}
\Big)
-\frac{2 e^{\nu_0}}{\beta_0 \pi}
\int_0^{\sqrt{s}/\Lambda}
dx \, x\sin(\nu_0 x)
\Big[ \frac{1}{x^2-\frac{1+\nu_0}{1-\nu_0}}  -  \frac{1}{x^2+1} \Big],
\end{equation}
for $s< (\frac{1+\nu_0}{1-\nu_0})\,\Lambda^2$, and

\begin{equation}\label{U1N1b}
\mathfrak{A}_1^{(N=1)}(s) =
\frac{1}{\beta_0}
\Big(
+\frac{1}{2}-\frac{1}{\pi}\arctan\frac{\log(s/\Lambda^2)}{\pi}
\Big)
+\frac{2 e^{\nu_0}}{\beta_0 \pi}
\int_{\sqrt{s}/\Lambda}^\infty
dx \, x\sin(\nu_0 x)
\Big[ \frac{1}{x^2-\frac{1+\nu_0}{1-\nu_0}}  -  \frac{1}{x^2+1} \Big],
\end{equation}
for $s> (\frac{1+\nu_0}{1-\nu_0})\,\Lambda^2$.
Both expressions are logarithmically divergent as $s\rightarrow (\frac{1+\nu_0}{1-\nu_0})\,\Lambda^2$.
The parametrized contribution is given by

\begin{equation}\label{U1Paraa}
\mathfrak{A}_1^{\text{(para)}}(s) =
h_1 - \frac{2 h_1}{\pi}
\int_0^{\sqrt{s}/\Lambda}
dx \, x\sin(\eta x)
\Big[\frac{1}{x^2}-\frac{1}{x^2-2}-4(\frac{1}{2}-h_2) \frac{1}{(x^2-2)^2} \Big],
\end{equation}
for $s< 2\Lambda^2$, and

\begin{equation}\label{U1Parab}
\mathfrak{A}_1^{\text{(para)}}(s) =
\frac{2 h_1}{\pi}
\int_{\sqrt{s}/\Lambda}^\infty
dx \, x\sin(\eta x)
\Big[\frac{1}{x^2}-\frac{1}{x^2-2}-4(\frac{1}{2}-h_2) \frac{1}{(x^2-2)^2} \Big],
\end{equation}
for $s> 2\Lambda^2$.
$\mathfrak{A}_1^{\text{(para)}}(s)$ behaves like $(s-2\Lambda^2)^{-1}$ for $s\approx 2\Lambda^2$.

\begin{figure}
 \centering\epsfig{file=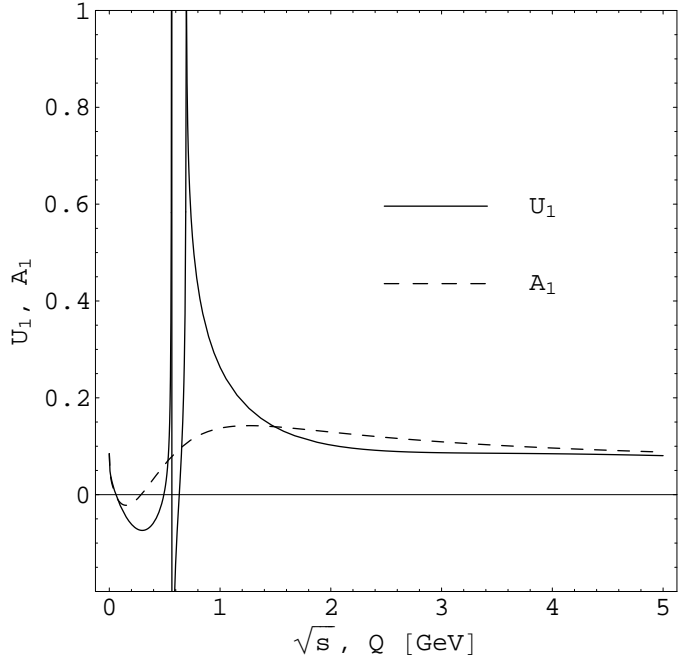}
\caption{\footnotesize The Minkowskian coupling $U_1^{(1)}(s)\equiv \mathfrak{A}_1^{(1)}(s)$
and Euclidean coupling $\mathcal{A}_1^{(1)}(Q^2)$ are plotted as a function of the energy scales $\sqrt{s}$ and $Q$, respectively.
The case $N=1$ is considered, the relevant expressions are given in Eq.\ (\ref{A1N1}) and Eqs.\ (\ref{U1N1a})-(\ref{U1Parab}).
The parameters are fixed as follows:
$\eta=0.3$, $h_1=0.1$, and $h_2=0$.
The value $\Lambda=0.4$ $GeV$ is used.}
\label{FigU1}
\end{figure}

In Fig.\ \ref{FigU1} the Minkowskian coupling $\mathfrak{A}_1$ is plotted together with
the Euclidean coupling $\mathcal{A}_1$ for a fixed set of parameters.
In the IR the both couplings differ drastically,
while for energies $\sim 2-3$ $GeV$ and higher
the difference is just the usual (perturbative) $\pi^2$-term \cite{Radyushkin:1982kg,Pivovarov:1991bi}.

The Minkowskian coupling can be obtained also for higher loop values $N$.
The resulting expressions are rather long and hence not shown here.
For resummations, it is usually easier to use the Euclidean coupling.

\section{Evaluation of Observables}

\subsection{Loop-level Dependence and RS Fixing}\label{SubSection:RSfixing}

For an energy scale relatively near to the boundary of the applicability region of pQCD,
let us say $Q=4$ $GeV$,
$a_\text{pt}^{(N=3)}$ and $a_\text{pt}^{(N=4)}$ differ by $\approx$ $2\%$,
while the difference between $a_\text{pt}^{(N)}$ and $\mathcal{A}_1^{(N)}$
is much smaller (see Figures \ref{FigA1ll234zoom} and \ref{FigA1examples}).
Thus, in the perturbative region $\mathcal{A}_1^{(N)}$ mimics $a_\text{pt}^{(N)}$ and
the well established methods of pQCD can be applied using the analytic coupling $\mathcal{A}_1^{(N)}$.

As the value of $Q$ decreases, the standard perturbative predictions for an observable $\mathcal{D}(Q^2)$ become unstable against variations in the number of loops and against the RS choice, and show a bad apparent convergence of the first terms of the series.
All this is related to the strong dependence of the low-scale perturbative coupling
under scale and scheme variations, and consequently also under the variation of $N$.
In turn, this is related to the vicinity of the Landau singularities.
For $Q=1$ $GeV$ the $N$-dependence of the coupling can be seen in Fig.\ \ref{FigA1ll234},
for the $\overline {\rm MS}$-scheme.
The couplings $a_\text{pt}^{(N=3)}$ and $a_\text{pt}^{(N=4)}$ differ by $\approx$ $40\%$.
In contrast, the difference between the analytical couplings
$\mathcal{A}_1^{(N=3)}$ and $\mathcal{A}_1^{(N=4)}$ is $\approx$ $5\%$.
In general, we shall call a coupling $\mathcal{A}_1^{(N)}$
{\it ``in the intermediate region $N$-independent coupling"}
if it varies at $Q\approx 1$ $GeV$ by $5\%$ or less when varying $N$ (with $N\geq 2$).

Let us evaluate a space-like observable $\mathcal{D}$ using truncated perturbation series at three- and four-loop-level,
and using the analytic coupling $\mathcal{A}_1^{(N)}$.
In this Subsection we analytize the non-leading terms of the series making the replacement:
$(a_\text{pt}^{(N)})^k \mapsto (\mathcal{A}_1^{(N)})^k$; however,
a different approach will be discussed in the next Subsection.
The three- and four-loop predictions for the leading twist contributions are:

\begin{eqnarray}
  \mathcal{D}^\text{(N=3)} &=& \mathcal{A}_1^{(3)}+ d_1(\mathcal{A}_1^{(3)})^2+ d_2(\mathcal{A}_1^{(3)})^3,\\
  \mathcal{D}^\text{(N=4)} &=& \mathcal{A}_1^{(4)}+ d_1(\mathcal{A}_1^{(4)})^2+ d_2(\mathcal{A}_1^{(4)})^3 + d_3(\mathcal{A}_1^{(4)})^4,
\label{Oll34}
\end{eqnarray}
where, of course, $\mathcal{D}^\text{(N)}$ and $\mathcal{A}_1^{(N)}$ are functions of $Q^2$.
In the intermediate energy region ($Q\sim 1$ $GeV$),
where $\mathcal{A}_1$ can be of order one,
powers of  $\mathcal{A}_1$ (in particular the term $d_3(\mathcal{A}_1^{(4)})^4$) are not necessarily suppressed with respect to $\mathcal{A}_1^{(4)}$.
Hence, given the experimental value of $\mathcal{D}(Q^2)$,
the extracted values of $\mathcal{A}_1^{(N)}$ will depend strongly on $N$.
This is problematic, because it is very unlikely to have $N$-stable observable predictions together with a $N$-unstable coupling.
A possible way to obtain an in the intermediate region $N$-independent coupling is to choose the RS separately for each observable
(as in the effective charge method \cite{ECH, KKP,Gupta}),
demanding for each observable: $d_3=0$.
Thus, extracting the coupling from a measured $\mathcal{D}$, we obtain from Eq.\ (\ref{Oll34})
$\mathcal{A}_1^{(3)}=\mathcal{A}_1^{(4)}$.
On the other hand, from Eq.\ (\ref{coupling})

\begin{equation}
\mathcal{A}_1^{(N=4)}-\mathcal{A}_1^{(N=3)}=
 \Big(\sum_{m=0}^{3} k_{4m}\,\log^m L_1 \Big)
\times\frac{1}{L_0^4}.
\label{couplingdifference}
\end{equation}
The latter difference is normally smaller than higher order uncertainties and experimental errors.
We conclude that an in the intermediate region $N$-independent coupling $\mathcal{A}_1^{(N)}$, used together with an observable-dependent RS choice,
provides a $N$-stable framework for the evaluation of observables at intermediate scales which is consistent within the errors.

Let us shortly review how within the evaluation method of Milton, Solovtsov, Solovtsova and Shirkov (MSSSh)
\cite{ShS,Milton:1997mi,Sh,Shirkov:2006gv}
the $N$-stability problem of observables at intermediates energies is solved.
In this approach, an observable in a given RS at $N$-loop-level is expressed as:

\begin{equation}
\mathcal{D}^\text{(N)}_{\text{MA}} = \mathcal{A}_{1,\text{MA}}^{(N)}+ d_1 \mathcal{A}_{2,\text{MA}}^{(N)}+ \ldots + d_{N-1} \mathcal{A}_{N,\text{MA}}^{(N)}.
\label{OMSSSh}
\end{equation}
The ``minimal analytic" (MA) functions $\mathcal{A}_{k,\text{MA}}^{(N)}(Q^2)$ \cite{ShS,Milton:1997mi,Sh,Shirkov:2006gv}
are computed using dispersion relations,
where the spectral functions are obtained from the functions $(a_\text{pt}^{(N)}(Q^2))^k$.
Hence, perturbative and MA couplings possess the same discontinuity along the Minkowskian semiaxis.
The behavior of $\mathcal{A}_{k,\text{MA}}^{(N)}(Q^2)$ for large $Q^2$ is the same as the one of $(a_\text{pt}^{(N)}(Q^2))^k$,
modulo (unwished) power-behaved terms.
The analytization procedure is unique, in the sense that it does not introduce any new parameter.
Thus, the IR behavior of the couplings is not modeled but given by the first coefficients of the $\beta$-function;
from our point of view, this a limitation of the approach.
An interesting feature of the MSSSh procedure, connected with the $N$-dependence of the coupling,
is that the couplings $\mathcal{A}_{k,\text{MA}}^{(N)}$ (with $k\geq 2$)
are suppressed compared to $(\mathcal{A}_{1,\text{MA}}^{(N)})^k$.
As a consequence, the analytized truncated perturbation series is rather stable against variations in the number of loops and against the RS choice.
It also shows well the apparent convergence of the first terms of the series.

\subsection{Skeleton-motivated Approach}\label{SubSection:skeleton}

In this Subsection we present the skeleton-motivated approach for the evaluation of observables \cite{Cvetic:2006mk,Cvetic:2006gc}.
In order to be specific, we consider the next-to-next-to-leading order case,
other approximations being generalizations of it.
The third order renormalization-group (RG) improved truncated perturbation series has the form:

\begin{equation}
     \mathcal{D}_{\text{pt}}(Q^2) = a_{\text{pt}}(Q^2)+ d_1\, a_{\text{pt}}^2(Q^2)+ d_2\, a_{\text{pt}}^3(Q^2),
     \label{PtTrunSeries}
\end{equation}
where the perturbative coupling $a_{\text{pt}}$ obeys the third-order RG equation:

\begin{equation}
\frac{\partial a_{\text{pt}}}{\partial\log Q^2}
= -[\,\beta_0\, a_{\text{pt}}^2(Q^2) +
\beta_1\, a_{\text{pt}}^3(Q^2)+\beta_2\, a_{\text{pt}}^4(Q^2)].
\label{rg}
\end{equation}

The skeleton expansion exists in QED if light-by-light subdiagrams are excluded \cite{Bjorken:1965}.
In QCD, its existence is not certain, but it can be postulated \cite{Gardi:1999dq,Brodsky:2000cr}.
The evaluation approach is motivated by the skeleton expansion but it does not rely on its existence.
The skeleton expansion is given by:

\begin{equation}
\mathcal{D}_{\text{skel}}(Q^2) =
\int_0^\infty \frac{dt}{t}\: F_{\mathcal{D}}^{\cal{E}}(t) \:
a_{\text{pt}}(t Q^2)
+ \sum_{n=2}^{\infty} s_{n-1}^\mathcal{D}
\left[ \prod_{j=1}^{n} \!\int_0^{\infty}\!\frac{d t_j}{t_j}
a_{\text{pt}}(t_j Q^2) \right]
F_{\mathcal{D}}^{\cal{E}}(t_1,\!\ldots\!,t_n),
\label{sk1}
\end{equation}
where the Euclidean {\it characteristic functions} $F_{\mathcal{D}}^{\cal{E}}(t_1,\!\ldots\!,t_n)$
are symmetric functions and are normalized as:

\begin{equation}
\int_0^\infty \frac{dt}{t} \ F_{\mathcal{D}}^{\cal{E}}(t)=1,
\qquad
\int \frac{dt_1}{t_1}\frac{dt_2}{t_2} \
F_{\mathcal{D}}^{\cal{E}}(t_1,t_2)=1, \  \ldots,
\label{mom0}
\end{equation}
and $s_i^\mathcal{D}$ are the skeleton coefficients.
It is assumed that the characteristic functions do not have any dependence on the running coupling constant.
If the representation (\ref{sk1}) exists,
its expansion in powers of $a_{\text{pt}}(Q^2)$ must correspond to the all order perturbation series of the considered observable.
The IR contributions of the integrals in Eq.\ (\ref{sk1}) suffer,
as we discussed earlier, from ambiguities due to the singularity and cuts of the perturbative coupling constant.

\begin{figure}
 \centering\epsfig{file=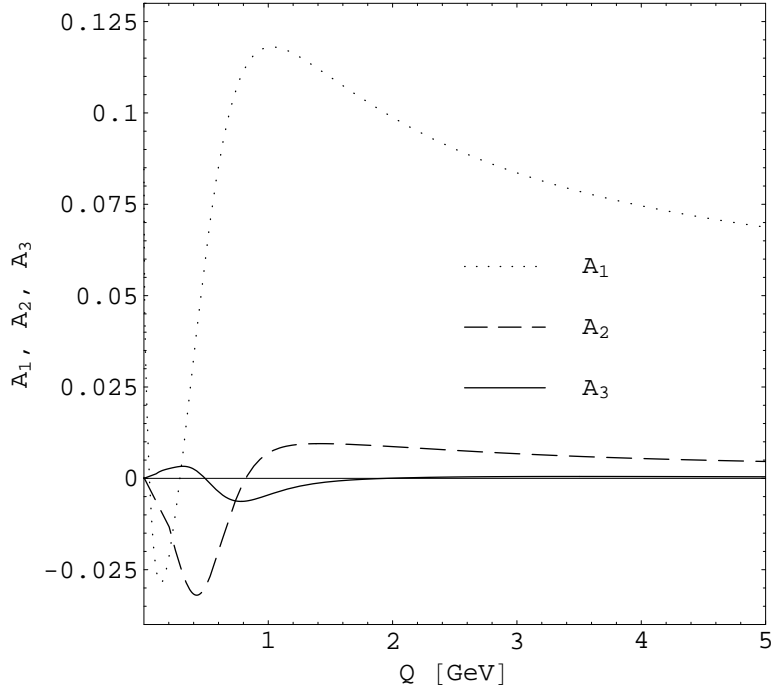}
\caption{\footnotesize The couplings $\mathcal{A}_2$ and $\mathcal{A}_3$,
defined in  Eqs.\ (\ref{rgAn1}) and (\ref{rgAn2}),
together with the corresponding coupling $\mathcal{A}_1$, defined in Subsection \ref{SubsectionNloopCoupling},
are plotted as a function of $Q$, in the $\overline {\rm MS}$-scheme, with $\Lambda=0.4$  $GeV$.
The parameters used for the couplings are $\eta = 0.3$, $h_1=0.1$, and  $h_2=0$.}
\label{FigA1A2A3}
\end{figure}

The analytization is formally made by replacing the perturbative coupling $a_{\text{pt}}(Q^2)$ of all skeleton integrands in Eq.\ (\ref{sk1}),
by an analytical coupling $\mathcal{A}_1(Q^2)$.
The replacement fixes the IR ambiguities of the integrals.
In order to obtain the analytic truncated series analogous to Eq.\ (\ref{PtTrunSeries}),
the function $\mathcal{A}_1(Q^2)$ is Taylor-expanded around (here, for simplicity) $t=1$ inside all skeleton terms.
In the considered third order treatment we need the first and second derivatives of $\mathcal{A}_1(Q^2)$.
Instead of the derivatives we use equivalently the analytical functions  $\mathcal{A}_2(Q^2)$ and  $\mathcal{A}_3(Q^2)$,
which are defined by:

\begin{eqnarray}
\frac{\partial \mathcal{A}_1}{\partial\log Q^2}
&=& -[\,\beta_0\, \mathcal{A}_2(Q^2) + \beta_1\, \mathcal{A}_3(Q^2)],\label{rgAn1} \\
\frac{\partial^2 \mathcal{A}_1}{\partial(\log Q^2)^2}
&=& 2\beta_0^2 \, \mathcal{A}_3(Q^2).
\label{rgAn2}
\end{eqnarray}
The latter equations are the truncated RG equation and the first derivative of it
(in higher order treatments, higher derivatives are involved),
with the replacement $a_{\text{pt}}^k \mapsto\mathcal{A}_k$.
Perturbatively, the differences
$\mathcal{A}_2 - a_{\text{pt}}^2$ and $\mathcal{A}_3 - a_{\text{pt}}^3$
are of $\mathcal{O}(a_{\text{pt}}^4)$.
In Fig.\ \ref{FigA1A2A3}, the couplings  $\mathcal{A}_2$ and $\mathcal{A}_3$ obtained using
the analytic coupling without power corrections defined in Subsection \ref{SubsectionNloopCoupling}
are plotted .

After Taylor-expanding, the terms corresponding to order $a_{\text{pt}}^4$ (e.g. $\mathcal{A}_2^2$ or $\mathcal{A}_1 \mathcal{A}_3$)
and higher are dropped and the skeleton integrals are formally performed.
As a result, we obtain:

\begin{eqnarray}
\mathcal{D}_{\text{an}}(Q^2)&=&
\mathcal{A}_1(Q^2)
+ \left[ \beta_0 f_1^\mathcal{D}(1) \mathcal{A}_2(Q^2)+
s_1^\mathcal{D}\mathcal{A}_1^2(Q^2) \right]
\label{AnTrunSeries}\\
&&
+ \left[
(\beta_0^2 f_2^\mathcal{D}(1)+ \beta_1 f_1^\mathcal{D}(1))
\mathcal{A}_3(Q^2)
+  2 s_1^\mathcal{D} \beta_0 f_{1,0}^\mathcal{D}(1)
\mathcal{A}_1(Q^2)\mathcal{A}_2(Q^2)
+ s_2^\mathcal{D} \mathcal{A}_1^3(Q^2) \right],
\nonumber
\end{eqnarray}
with the momenta
\begin{equation}
\begin{split}
f_i^\mathcal{D}(t_*)
&=
\int_0^\infty \frac{dt}{t}
  F_{\mathcal{D}}^{\cal{E}}(t)(-\log t/t_*)^{i},
\\
f_{i,j}^\mathcal{D}(t_*)
&=
\int \frac{dt_1}{t_1}\frac{dt_2}{t_2}
  F_{\mathcal{D}}^{\cal{E}}(t_1,t_2)
  (-\log t_1/t_*)^{i}(-\log t_2/t_*)^{j}.
\end{split}
\label{mom1}
\end{equation}
Thus, comparing Eqs.\ (\ref{PtTrunSeries}) and  (\ref{AnTrunSeries}),
we see that in the analytization process
$a_{\text{pt}}$ is replaced by $\mathcal{A}_1$,
$a_{\text{pt}}^2$ is replaced by a linear combination of $\mathcal{A}_2$ and $\mathcal{A}_1^2$, and
$a_{\text{pt}}^3$ is replaced by a linear combination of $\mathcal{A}_3$, $\mathcal{A}_1 \mathcal{A}_2$, and $\mathcal{A}_1^3$.
In the UV regime, the difference between both equations is of order $a_{\text{pt}}^4$.
The coefficients $s_1^\mathcal{D}$ and  $s_2^\mathcal{D}$ and the values of
$f_1^\mathcal{D}$, $f_2^\mathcal{D}$ and $f_{1,0}^\mathcal{D}$ can be obtained in a given RS,
for the third order treatment,
from the perturbative coefficients $d_1$ and $d_2$
using their dependence on the number of flavors $N_f$ (see \cite{Cvetic:2006mk,Cvetic:2006gc}).

The definitions (\ref{rgAn1}) and (\ref{rgAn2}) are made in such a way that,
the result of the third order Taylor-expansion of the analytic coupling inside the one-chain term

\begin{equation}
\int_0^\infty \frac{dt}{t}\: F_{\mathcal{D}}^{\cal{E}}(t) \:
\mathcal{A}_1(t Q^2)
\rightarrow
\mathcal{A}_1(Q^2)
+  \beta_0 f_1^\mathcal{D}(1) \mathcal{A}_2(Q^2)
+ (\beta_0^2 f_2^\mathcal{D}(1)+ \beta_1 f_1^\mathcal{D}(1)) \mathcal{A}_3(Q^2),
\end{equation}
corresponds to the perturbative third order truncated expression

\begin{equation}
\int_0^\infty \frac{dt}{t}\: F_{\mathcal{D}}^{\cal{E}}(t) \:
a_{\text{pt}}(t Q^2)
=
a_{\text{pt}}(Q^2)
+  \beta_0 f_1^\mathcal{D}(1) a_{\text{pt}}^2(Q^2)
+ (\beta_0^2 f_2^\mathcal{D}(1)+ \beta_1 f_1^\mathcal{D}(1)) a_{\text{pt}}^3(Q^2)
+\mathcal{O}(a_{\text{pt}}^4),
\end{equation}
with $a_{\rm pt}^k \mapsto \mathcal{A}_k$.
A different choice of
$\mathcal{A}_2$ and $\mathcal{A}_3$
would not allow this correspondence.

If the exponentially modified coupling of Section \ref{Section:ExpCoup} is used in Eq.\ (\ref{AnTrunSeries}),
no power-behaved terms are introduced.

On the other hand, if the characteristic function $F_{\mathcal{D}}^{\cal{E}}(t)$ is known,
the object we are interested in is the analytic version of the one-chain resummed extension of Eq.\ (\ref{PtTrunSeries}).
In order to obtain it we expand the skeleton integrals as before,
except for the leading term which we keep unexpanded. We obtain

\begin{equation}
\mathcal{D}_{\text{res,an}}(Q^2) =
\int_0^\infty \frac{dt}{t}\: F_{\mathcal{D}}^{\cal{E}}(t) \:
\mathcal{A}_1(t Q^2) +
s_1^\mathcal{D} \mathcal{A}_1^2(Q^2)
+ \left[ s_2^\mathcal{D} \mathcal{A}_1^3(Q^2)
+ 2 s_1^\mathcal{D} \beta_0 f_{1,0}^\mathcal{D}(1)
\mathcal{A}_1(Q^2)\mathcal{A}_2(Q^2)
\right].
\label{sk2}
\end{equation}
Besides, as we motivated in the previous Subsection,
in order to decrease the $N$-dependence of the coupling, we adopt
an observable-dependent RS choice,
such that each observable is written as

\begin{equation}
\mathcal{D}(Q^2) =
\int_0^\infty \frac{dt}{t}\: F_{\mathcal{D}}^{\cal{E}}(t) \:
\mathcal{A}_1(t Q^2)
+ s_1^\mathcal{D} \mathcal{A}_1^2(Q^2) +
\mathcal{O}_n,
\label{v1}
\end{equation}
i.e., evaluation in a RS where $s_2^\mathcal{D}$ and $f_{1,0}^\mathcal{D}$ are equal to zero
(again, see \cite{Cvetic:2006mk,Cvetic:2006gc} for details).
The error, for the present (third order) case, is formally of order four, $\mathcal{O}_n=\mathcal{O}_4$.
The final expression Eq.\ (\ref{v1}) remains unchanged in higher order treatments,
the error $\mathcal{O}_n$  being one order of magnitude higher than the order of the perturbative expression.

This is one variant of the evaluation approach.
Other possibilities can be found in \cite{Cvetic:2006mk,Cvetic:2006gc}.
In particular, a different RS can be chosen \cite{Cvetic:2006gc} such that the second term of Eq.\ (\ref{v1})
is replaced by the first derivative of the analytic coupling

\begin{equation}
\mathcal{D}^{'}(Q^2) =
\int_0^\infty \frac{dt}{t}\: F_{\mathcal{D}}^{\cal{E}}(t) \:
\mathcal{A}_1(t Q^2)
+ \tilde{s}_1^\mathcal{D} \frac{\partial \mathcal{A}_1}{\partial\log Q^2} +
\mathcal{O}_n,
\label{var}
\end{equation}
where the derivative of $\mathcal{A}_1$ can be written in terms of $\mathcal{A}_2$ and $\mathcal{A}_3$
according to Eqs.\  (\ref{rgAn1}) and (\ref{rgAn2}).
In Refs. \cite{Cvetic:2006mk,Cvetic:2006gc} this evaluation approach is applied
to the Adler function, the Bjorken polarized sum rule and
the semihadronic $\tau$ decay ratio,
using three different models for $\mathcal{A}_1$ (which do introduce running coupling power-terms).

\section{Conclusion}

Motivated by the belief that relevant non-perturbative QCD contributions arise only from the integration of IR degrees of freedom,
a new class of models for an analytic QCD coupling is proposed.
The main characteristic of the proposed analytic coupling is that its difference to the perturbative one, in the UV region,
is smaller than any inverse power of the energy.
As a consequence, a general analytization procedure of observables
which makes use of this coupling
yields predictions which have no power-behaved contributions arising from UV degrees of freedom.
In addition, the constructed coupling is finite when $Q\rightarrow 0$.
The analytic coupling is conceived as the central object
in the construction of observables with the required analytic properties.
A particular procedure is presented in Subsection \ref{SubSection:skeleton},
the skeleton-motivated approach for the evaluation of observables, developed in  \cite{Cvetic:2006mk,Cvetic:2006gc}.\\

For a coupling whose difference to the perturbative coupling
is a very suppressed power term $\sim (\Lambda^2/Q^2)^{k_{\rm max}}$ ($k_{\rm max}\gg 1$),
it is shown that in the one-chain resummation of space-like observables  $\mathcal{D}(Q^2)$,
the UV regime still contributes an appreciable power correction $\sim (\Lambda^2/Q^2)^n$.
Here $z=n$ is the location of the leading IR renormalon of $\mathcal{D}$ in the Borel plane,
and therefore, the leading OPE contribution to this observable is also $\sim (\Lambda^2/Q^2)^n$.
Thus, even the suppressed difference $\delta \mathcal{A}_1(Q^2)\sim (\Lambda^2/Q^2)^{k_{\rm max}}$
cannot be accommodated within the ITEP interpretation of the OPE,
which says that the power-suppressed terms in $\mathcal{D}(Q^2)$ are contributions from the IR regime only.
We emphasize that the last conclusion is obtained using the criterium for the identification of UV contributions,
in the one-chain resummation integral, described in Section \ref{Section:PowCorr}.
It is based on comparing the momentum flowing through the photon/gluon internal propagator,
with the characteristic scale of the strong interactions, $\Lambda$.\\

An analytization procedure that makes use of the proposed coupling,
must be considered as a phenomenological attempt with the main motivation to
enlarge the energy range of applicability, and improve the description capability and the predictive power
of perturbative QCD.
Of course, we are most interested in the intermediate energy region $Q\approx 1-2$ $GeV$.
The model includes, in addition to the QCD scale, a number of parameters (three parameters)
describing the IR behavior of the coupling.
In the intermediate energy region,
the proposed coupling has low $N$ (loop-level) and RS dependence.
By comparing predictions with experimental data for low-energy observables,
the three parameters can in principle be fixed and the low-energy behavior of the coupling can be obtained.
If, in addition, an all order resummation is made, as in Eqs.\ (\ref{v1}) or (\ref{var}),
then values of the coupling at low $Q$ are ``probed". \\

In Subsection \ref{SubSection:RSfixing} we motivate an observable-dependent RS-fixing criterium.
It provides a consistent $N$-stable framework for the evaluation of observables,
using a coupling with low $N$-dependence at intermediate scales.\\

The evaluation approach given by Eq.\ (\ref{v1}) or by Eq.\ (\ref{var})
generates power-behaved terms only of IR nature.
However, it is likely that the model cannot describe all non-perturbative effects relevant to low-energy observables.
For example, chiral symmetry breaking effects are probably not accounted for within this framework.
This motivates a mixed approach.
Namely, the one where to  Eq.\ (\ref{v1}) or (\ref{var}) OPE power-behaved terms are added.
As in the standard one-chain resummation,
the expectation is that non-perturbative contributions generated by the analytic coupling correspond to
an important part  of the non-perturbative effects
otherwise carried by OPE terms
in the standard (perturbative QCD plus  OPE) approach.

\section*{Acknowledgments}
The authors acknowledge helpful discussions with M.~Beneke.
This work was supported in part by Fondecyt grant 1050512 (G.C.)
and Conicyt Fellowship 3060106 (C.V.).

\end{document}